\documentclass[a4paper,twocolumn,fleqn,usenatbib]{mnras}

\usepackage[T1]{fontenc}
\usepackage{ae,aecompl}
\usepackage[none]{hyphenat} 

\usepackage[]{graphicx}

\usepackage{float}
\usepackage{amsmath}	
\usepackage{amssymb}	
\usepackage{textcomp}
\usepackage{mathptmx}

\usepackage{caption}
\usepackage{subcaption}
\usepackage{multicol}

\title[]{Measuring Electric Dipole Moments of Trapped Sub-mm Particles}

\author[Onyeagusi et al.]{
F. Chioma Onyeagusi\thanks{E-mail: florence.onyeagusi@uni-due.de},
Jens Teiser, Niclas Schneider,
Gerhard Wurm\\
University of Duisburg-Essen, Faculty of Physics, Lotharstr. 1, 47057 Duisburg, Germany
}

\date{}

\pubyear{2021}

\begin{document}
\label{firstpage}
\pagerange{\pageref{firstpage}--\pageref{lastpage}}
\maketitle

\begin{abstract}
We present a method for measurements of electric dipole moments on (sub)-mm size (basalt) particles levitated in an acoustic trap and centered within a plate capacitor. If an electric field is applied the particles oscillate with specific frequencies due to their permanent dipole moments. We observe dipole moments on the order of $D_P = 10^{-15} ... 10^{-14} \rm \, C \, m $. The dipole moment increases in small aggregates with the number of grains and is larger for samples vibrated (tribocharged) before trapping. The basalt grains show no sign of change in their dipole moment during measurements, implying a timescale for charge mobility being at least larger than minutes.
\end{abstract}

\begin{keywords}
electrical dipole moment, granular medium, charging
\end{keywords}



\section{Introduction}

Tribocharging is ubiquitous in granular matter \citep{Lacks2019, Harper2018, Lee2015, Forward2009, Haeberle2018, Mccarty2007}. Whenever two particles collide they exchange charge, whether they are made of metal or a dielectric. For charge transfer at the contact various mechanisms  might be at work. The discussed mechanisms range from electrons being exchanged from one trapped energy state on one particle to some trapped state on the other, over material exchange, to volatiles like water being ionized and transferred \citep{Lee2018, Lacks2011, Pan2019}. There are well-known biases, e.g. small grains are usually charged negatively in collisions with large grains \citep{Forward2009b, Waitukaitis2014}. In this case, subsequent size segregation by sedimentation will lead to large-scale charge separation. This could eventually produce electric fields strong enough for atmospheric breakdown, e.g. notable as thunderstorms or lightning during volcanic eruptions \citep{Cimarelli2014, Harrison2016}. Conditions for such breakdowns might be reached on particle distances already \citep{Matsuyama2018, Wurm2019}. Radiation is also detectable on smaller scales \citep{Wurm2019, Harper2021, Schoenau2021}. 

Generally, charges simply add a strong force field to a granular sample. This can promote the growth of stable „ionic“ particle aggregates which is an important aspect of planet formation \citep{Lee2015,Steinpilz2020a, Jungmann2021, Teiser2021}. Charges and related fields can also change sand transport behavior during aeolian events \citep{Kok_and_Renno_2006, Zhang_et_al_2015}. 

The dynamics of a charged system are often described in terms of net charge. However, charges produced by collisions on an insulator are not homogeneously distributed on the particle surface \citep{Baytekin,Grosjean,Steinpilz2020b}. Therefore, the electric fields are not only monopole fields but contain higher orders. Due to the distance dependence of the Coulomb forces, higher orders might be quite significant for interparticle attraction and repulsion if two particles are close to each other. When two patches of opposite polarity are close to contact, their corresponding grains might be attracted even if the overall net charges were of the same polarity \citep{Steinpilz2020a}. Complex charge patterns can be sampled by Kelvin Probe microscopy but not for free moving particles. Adding to the complexity are multipoles induced by external fields or the fields of other grains, which can also outweigh net charges \citep{Matias2018}. 

Here, we focus on permanent electric dipoles on freely levitated grains, to compare measured dipole moments with models of surface charge distributions resulting from collisions. This will help to further constrain existing models \citep{Steinpilz2020b, Grosjean}. Also, charge transport in and on dielectric grains and the timescales for charge redistribution can be studied this way. 
It, for example, has been considered that (sand) particles might readily accumulate charge if exposed to an external electric field, which changes the conditions that allow particles to be lifted from the ground \citep{Rasmussen2009, Kok_and_Renno_2006, Paehtz_et_al_2020}.
Timescales are yet unknown. However, the change of the charge distribution on a grain can easily be deduced if the temporal evolution of permanent dipole moments on grains is measured. This is also important for the stability of charged particle aggregates when and how charges rearrange over time, e.g. in the context of planet formation \citep{Steinpilz2020a}. In this context, dipole measurements can provide an initial indication on changes over time.

Dipoles, however, cannot be measured as easily as net charges by means of a Faraday cup. Charges can also be determined by analyzing a particle’s motion in an electric field. In microgravity, even small charges can be deduced from accelerations in a homogeneous electric field \citep{Jungmann2018, Teiser2021, Jungmann2021conserve, Jungmann2021}. This also offers a way to determine dipole moments as non-symmetric grains now visually oscillate in the electric field \citep{Steinpilz2020b}. It must  be kept in mind that any collision with a wall, which is inevitable even under microgravity, will change the charge state of a grain, eventually. Therefore, observations of individual grains trapped over an extended period of time are desirable. 

Paul traps are an option for smaller grains \citep{Schlemmer2001}. This kind of trapping, however, is sensitive to the charge, so large variations are harder to follow. One method that allows particles to be trapped regardless of their charge is acoustic levitation. \citet{Kline2020} analyzed the oscillations caused by net charge in an electric field for an acoustically trapped particle. Here, even large changes in net charge, e.g. transferred in a collision with a wall, do not change the trapping efficiency. 
We essentially follow this concept and also report experiments with grains in an acoustic trap and an electric field. However, we are rather interested in the rotational motion. So while \citet{Kline2020} study the linear oscillations of grains due to their net charge in the electric field of a capacitor, we study rotational motion, as the electric field induces oscillations with characteristic frequencies trying to align electric dipoles.

\section{Experiments\label{experiment}}

\begin{figure}
	\includegraphics[width=\columnwidth]{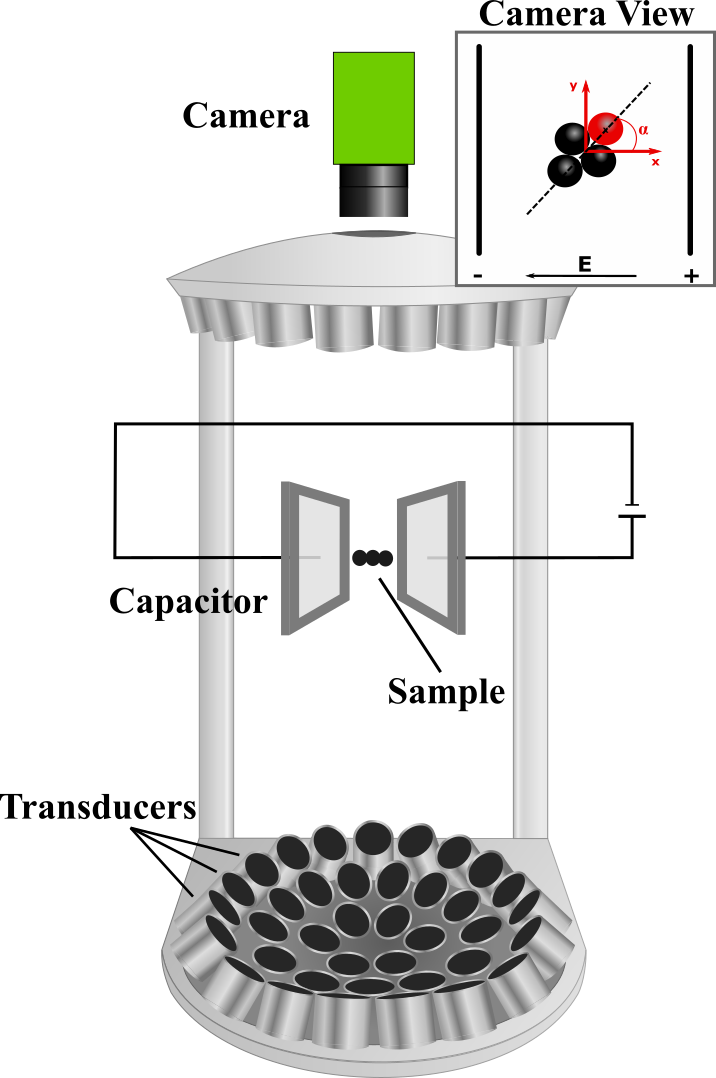}
	\caption{\label{fig.setup}Sketch of the setup; the basic acoustic levitator is a rebuild of a work by \citet{doi:10.1063/1.4989995}. The capacitor electrode meshes are 15 mm apart. A voltage up to 975 V is applied. To quantify the rotation of the particle cluster, the center of mass is tracked, together with one distinct particle. This particle is marked in red here and in table \ref{table} for all configurations. Not shown here, the whole setup is placed under an acrylic cover reducing convection.}
\end{figure}

A sketch of the experiment is shown in fig. \ref{fig.setup}. It shows the acoustic trap with its transducers operating at 40 kHz and a particle cluster which is observed from above. The capacitor currently consists of two steel meshes used as electrodes. We chose this kind of mesh as it is still acoustically transparent, which makes it less inclined to disturb the trapping field, yet fine enough to create an approximately homogeneous electric field in the center of the capacitor.

 Levitated aggregates always align in a plane perpendicular to the acoustic standing wave, i.e. are only extended in the horizontal x-y-plane and not in the vertical z-direction. Therefore, from the top view, the center of mass of the grains as well as specific individual grains can be traced, as they remain in the focal plane of the optical system. Particles are automatically tracked with Fiji \citep{fiji2012} using the plugin Trackmate \citep{Trackmate2017}. To parameterize the alignment of an agglomerate, we define the alignment angle $\alpha$ as sketched in fig. \ref{fig.setup} top right. 


\section{Particle rotation}

\subsection{Trap induced rotation}

For the present study, we use basalt spheres with a diameter of {0.9 and 0.55 mm}.
The grains are not motionless when captured in the acoustic trap. Regarding the dipoles, the rotational motion is of interest here. In the vertical direction, gravity is accounted for by the trap as it pushes the particles toward the pressure nodes of the standing wave. Small deviations from the equilibrium point lead to linear vertical motion. The not perfectly symmetric acoustic field can also lead to torques around one of the two horizontal axes. However, the amplitudes prove to be rather small compared to rotation around the vertical axis, so we do not consider this kind of motion any further.

\begin{figure}
    \centering
	\includegraphics[width=\columnwidth]{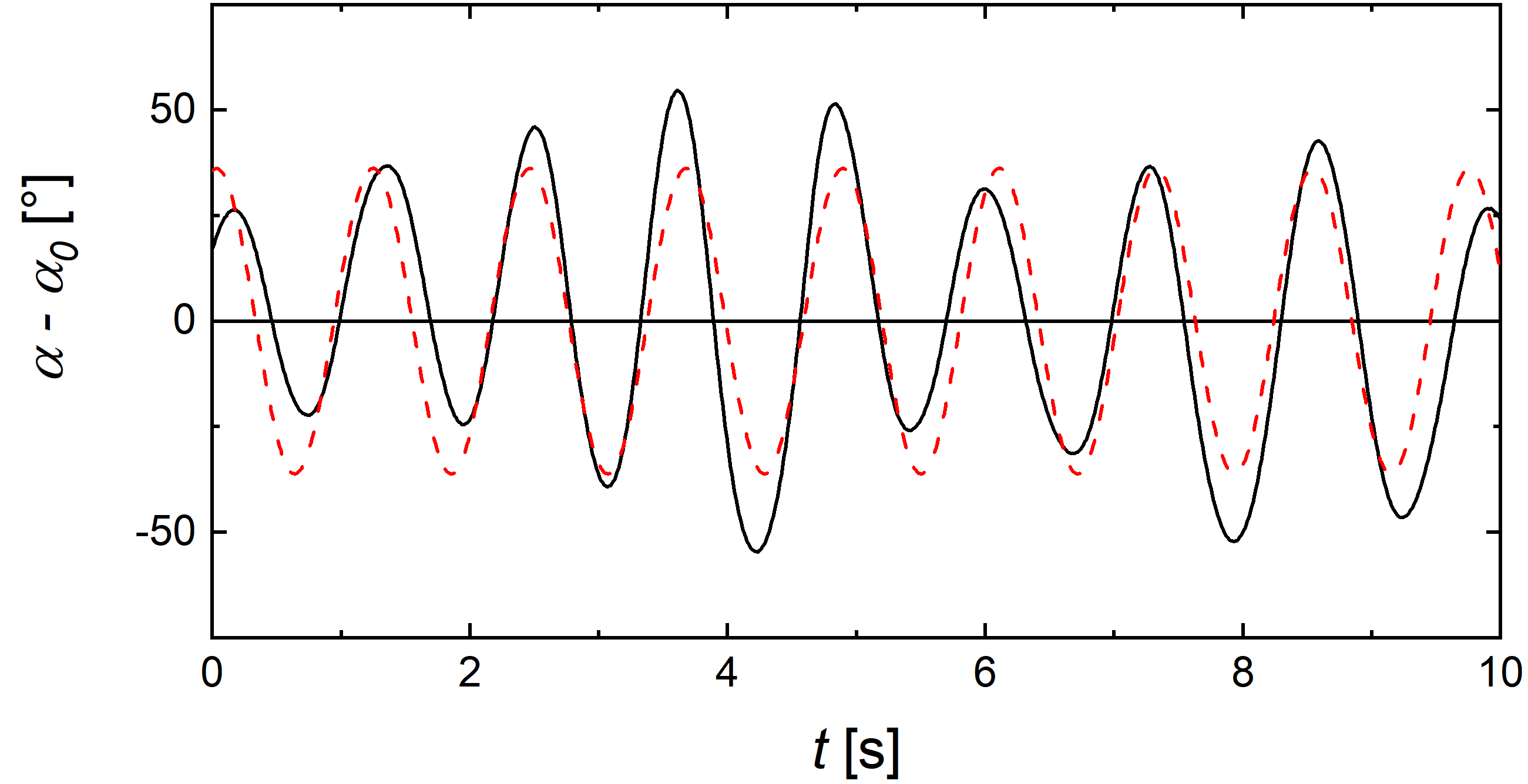}
	\caption{\label{fig.nofieldrotation}Oscillating rotational motion of a cluster of 2 grains around its vertical axis (solid line). To obtain the angle $\alpha$ we tracked the position of one particle (later in table \ref{table} marked in red) with regard to the center of mass of the cluster which is also determined. See also fig. \ref{fig.setup} top right inset of camera view for angle definition. The red dashed line is a harmonic fit to the data.}
\end{figure}

Within the horizontal plane, particles react more easily to disturbances. The attraction of net charges in the capacitor should deflect the grains from their equilibrium position, but this will not be the focus of our study. However, rotation around the vertical axis is frequently observed even without applying an electric field, especially as oscillations around a certain equilibrium angle $\alpha_0$ as seen in fig. \ref{fig.nofieldrotation}. 

There are random, small disturbances due to asymmetries of the trapping field and likely fluctuation of residual convection, but the motion can still be approximated as harmonic oscillation as seen in fig. \ref{fig.nofieldrotation}. At least within the scope of this paper, we consider the accuracy as sufficient as we are mostly interested in the oscillation frequency. A harmonic oscillation implies that a torque induced by the trap can be described as
\begin{equation}
M_T = C_T (\alpha-\alpha_T)\,\,.
\label{traptorque}
\end{equation}
Here,~$C_T$ is a constant, which depends on the trap, the specific structure of the cluster, and slightly on the electric field (see below). $\alpha$ is the angle of the elongation axis with respect to the equilibrium angle $\alpha_T$, which is specific to the torques of the trap. 

In general, the oscillation of a trapped cluster can be described by the equation of motion,
\begin{equation}
M = J \, \frac{d^2\alpha}{dt^2}
\label{oscillate}
\end{equation}
where $M$ is the torque and $J$ is the moment of inertia around the rotational axis through the center of mass. The aggregates considered are simple, flat aggregates of identical spherical particles, which oscillate only around the vertical axis. In this case, $J$ can be calculated from the moment of inertia for spheres using Steiner's law.

If the trap is active without an electric field, the above torque $M_t$ has to be used and the cluster rotates with an angular frequency

\begin{equation}
\omega = \sqrt{\frac{C_T}{J}}\,\,.
\label{omega_trap}
\end{equation}

\subsection{\textit{E}-field induced oscillation}

If an electric field $E$ is additionally applied in the horizontal direction, two further torques $M_E$ around the vertical axis are added. The first electric torque $M_E$ is the one induced by permanent charges sitting on the grains in random positions. 
For a discrete number of charges on the surface of the aggregate the permanent dipole moment $D_P$ is defined as the sum of all $N$ (discrete) charges $Q_i$ times the distances $R_i$ from the center of mass
\begin{equation}
{D_P} = \left| \sum_{i = 1}^N Q_i \vec{R}_i \right| \,\,.
\end{equation}
This dipole moment is constant and independent of the electric field for immobile charges.
Furthermore, it can be much larger than net charges placed on extreme positions might suggest \citep{Steinpilz2020b}. At its extreme, even grains with zero net charge can have large permanent dipole moments.

The torque according to this dipole is proportional to the electric field $E$ and described by
\begin{equation}
M_E = D_P \, E\, \sin (\alpha - \alpha_P) \,\,,
\label{oscillate2}
\end{equation}
where $\alpha_P$ is the respective equilibrium angle, which will usually not coincide with $\alpha_T$ of the acoustic trap. 
The electric field is given by $E=U/d$, with the capacitor voltage $U$ and the distance $d$ between the capacitor electrodes. 

A second potential source for dipoles is the electric field (see below).
In a symmetric particle like a single sphere, these induced dipoles are always aligned to the field. Independent of its field strength, such dipoles do not generate a torque as they will always rearrange upon rotation. For non-symmetric clusters, however, the induced dipole field is also non-symmetric and the external electric field produces a dipole component $D_i$ that induces a torque. As the induced dipole is proportional to the electric field, the torque would introduce an $E^2$ dependency in our data. Eventually, it would dominate in very strong fields. However, the applicable electric field is limited to values below the atmospheric breakdown level. For a dimer of 0.5 mm glass spheres, the induced torque was estimated by \citet{Steinpilz2020b} to be significantly smaller than the permanent dipole even at a large field of $E > 80 \mathrm{kV/m}$, or on the order of $D_i \sim 10^{-17} \rm Cm$ at maximum. While this might be larger for particles twice the size, this is still orders of magnitude smaller than the permanent dipoles quantified below ($\sim 10^{-14} \rm Cm$) in agreement with \citet{Steinpilz2020b}. Therefore, we do not consider the induced component any further.

\begin{figure}
    \centering
	\includegraphics[width=\columnwidth]{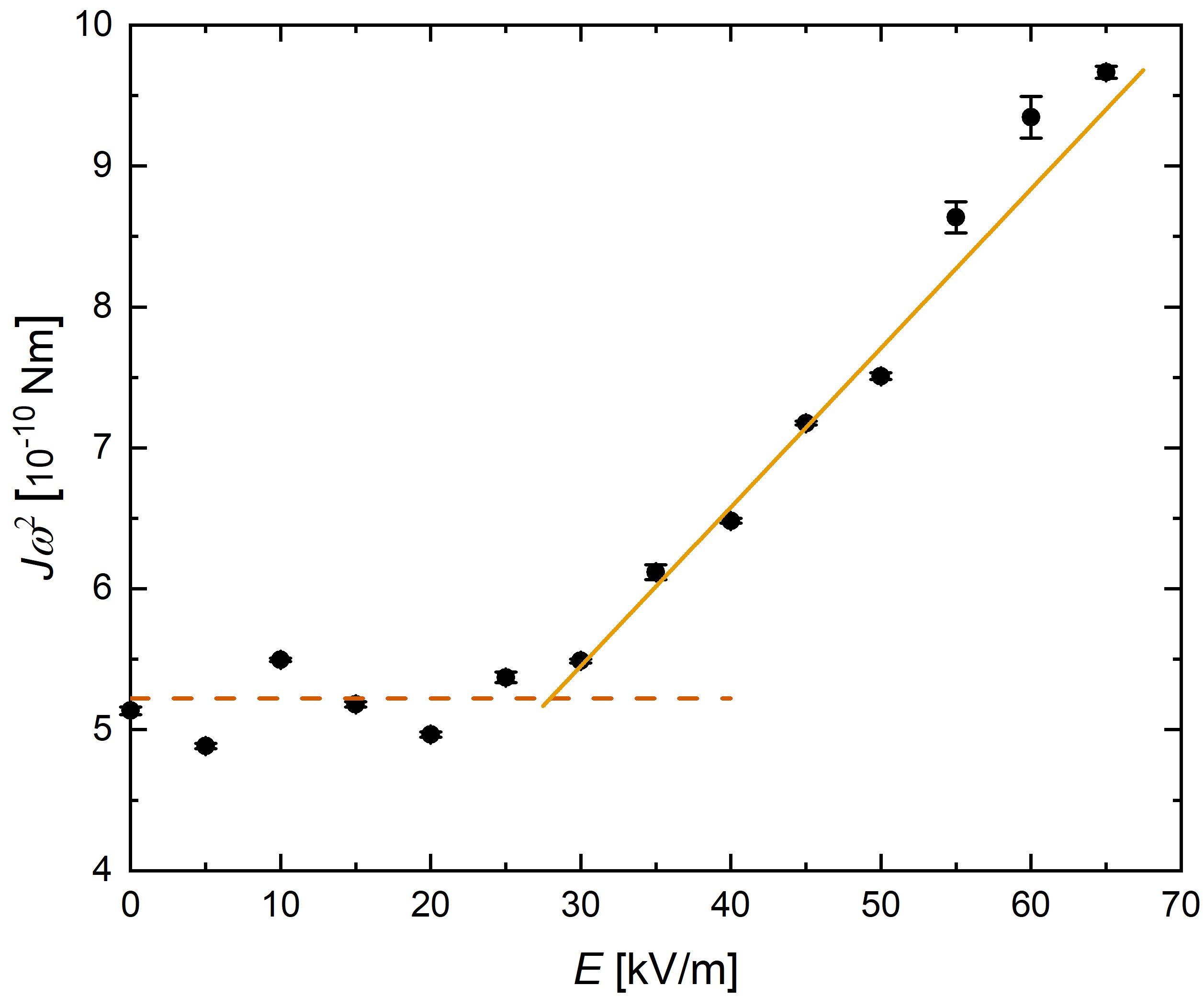}
	\caption{\label{fig.thatsit}Observed angular frequencies over the electric field (black dots) for a 6 particle aggregate; The solid line is a linear fit according to eq. \ref{eq.J}; The dashed line is a constant to visualize the region dominated by the trap which is not included in the fit.}
\end{figure}

Including the $E$-dependence of the total torque applied to the cluster, $M = M_T + M_E$, Eq. \ref{omega_trap} converts to
\begin{equation}
    J \omega^2 = C_T(E) + D_P\cdot E  
    \label{eq.J}
\end{equation}
where we approximate the sin function in eq. \ref{oscillate2} to be linear. 
At this point, a potential $E$-dependence of the acoustic trap has to be considered. This might be caused by a net charge shifting the particle slightly in the capacitor field, changing the behavior of the acoustic field. The effect can be seen in the example shown in fig. \ref{fig.thatsit}. 
For a constant $C_T$, one would expect a simple linear trend of $J \omega^2$. Indeed the data is linear at larger fields. However, it is constant at small field strengths. Evidently, the acoustic trap essentially cancels the linear increase up to a certain threshold. This behavior can be seen for all data. The threshold, where the linear dependence takes over depends on the absolute value of $C_T(0)$ and occurs for larger fields with increasing values of $C_T(0)$. In any case, we take this as empirical fact and fit eq. \ref{eq.J} only to the large field data as seen in fig. \ref{fig.thatsit}. 
For the depicted example, the resulting dipole moment is $D_P = 11.284 \times 10^{-15} \pm 0.489 \times 10^{-15}\rm Cm $.
The permanent dipole moments of all data are given in table \ref{table}.

\begin{figure}
    \centering
	\includegraphics[width=\columnwidth]{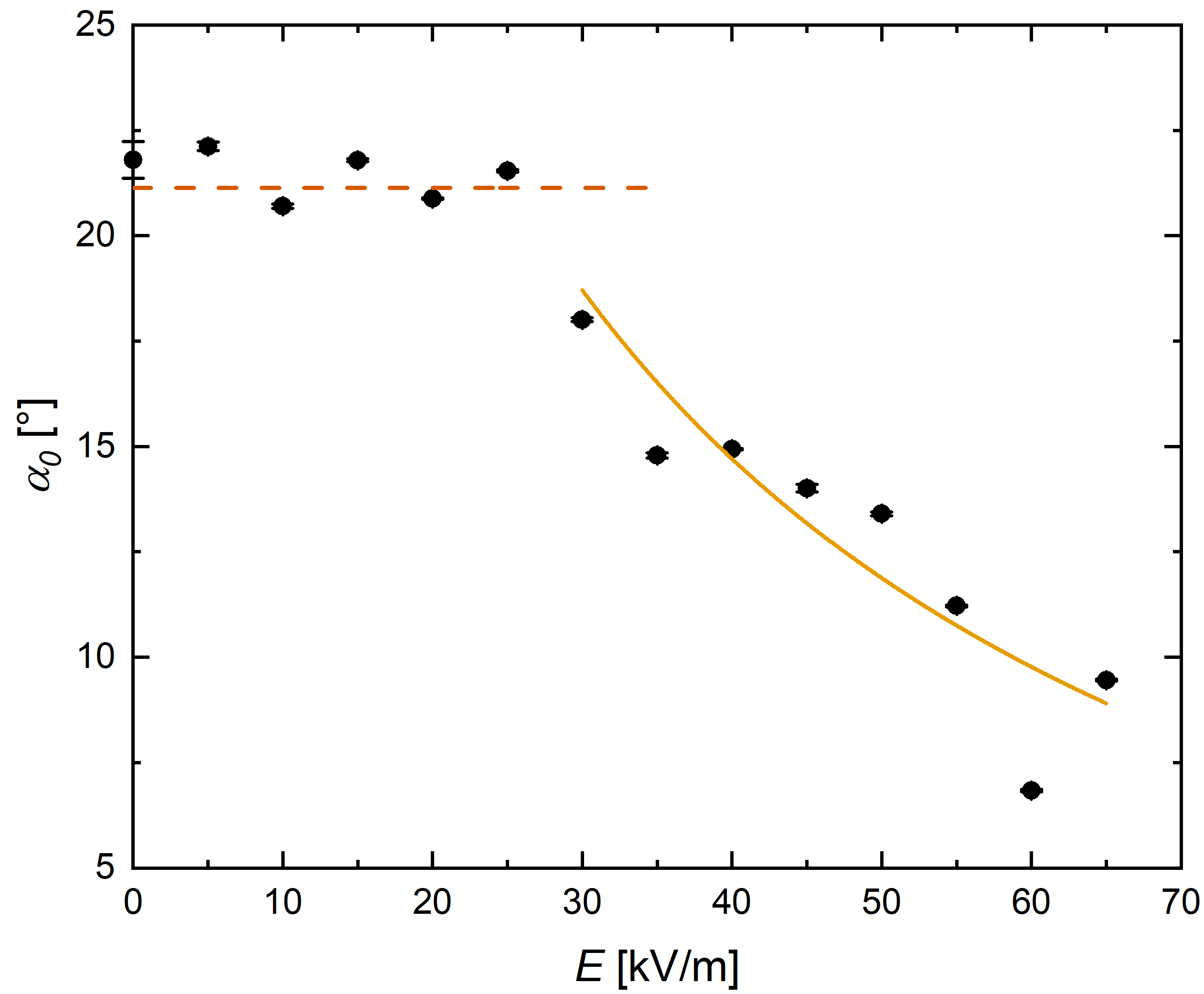}
	\caption{\label{fig.alignment}Shift of the orientation of the alignment from trap torque dominated to $E$-field torque dominated as the $E$-field increases. 
	Data for basalt (black dots); approximated as solid line is a fit according to eq. \ref{oscillate5} in the region where the field-induced torque starts to become visible for the same 6 particle aggregate as in fig. \ref{fig.thatsit}; The dashed line is a constant to guide the eye on the data points dominated by the trap.}
\end{figure}

Further information is provided by the equilibrium positions $\alpha_0$.
As mentioned before, depending on the cluster symmetry, the alignment of the cluster without $E$-field $\alpha_T$ will usually not coincide with the alignment of the cluster if an electric field is applied. For small electric fields, the torques of the trap dominate. If the $E$-field increases sufficiently, $M_E$ will dominate, eventually. Therefore, the equilibrium angle $\alpha_0$ will shift. 
In detail, in equilibrium $\alpha_0$ is described by
\begin{equation}
C_T (\alpha_0-\alpha_T) + D_P \, (\alpha_0 - \alpha_P) 
= 0
\label{oscillate4}
\end{equation}
or
\begin{equation}
\alpha_0 = \frac{C_T(E) \alpha_T + E D_P \alpha_P  
}{C_T(E)+E D_P 
} \,\,.
\label{oscillate5}
\end{equation}
This shift is also observed as shown in fig. \ref{fig.alignment}.
Here, approximating the data with eq. \ref{oscillate5} results in an equilibrium angle $\alpha_P = -4.630^{\circ} \pm 2.700^{\circ}$.
It has to be considered again, that $C_T$ initially depends on $E$ up to a certain value and only then can be considered constant with respect to the electric field. 
The agreement of the alignment with expected behavior is essentially a consistency check for the assumed model of two torques here.

\section{Discussion}

\begin{figure}
    \centering
	\includegraphics[width=\columnwidth]{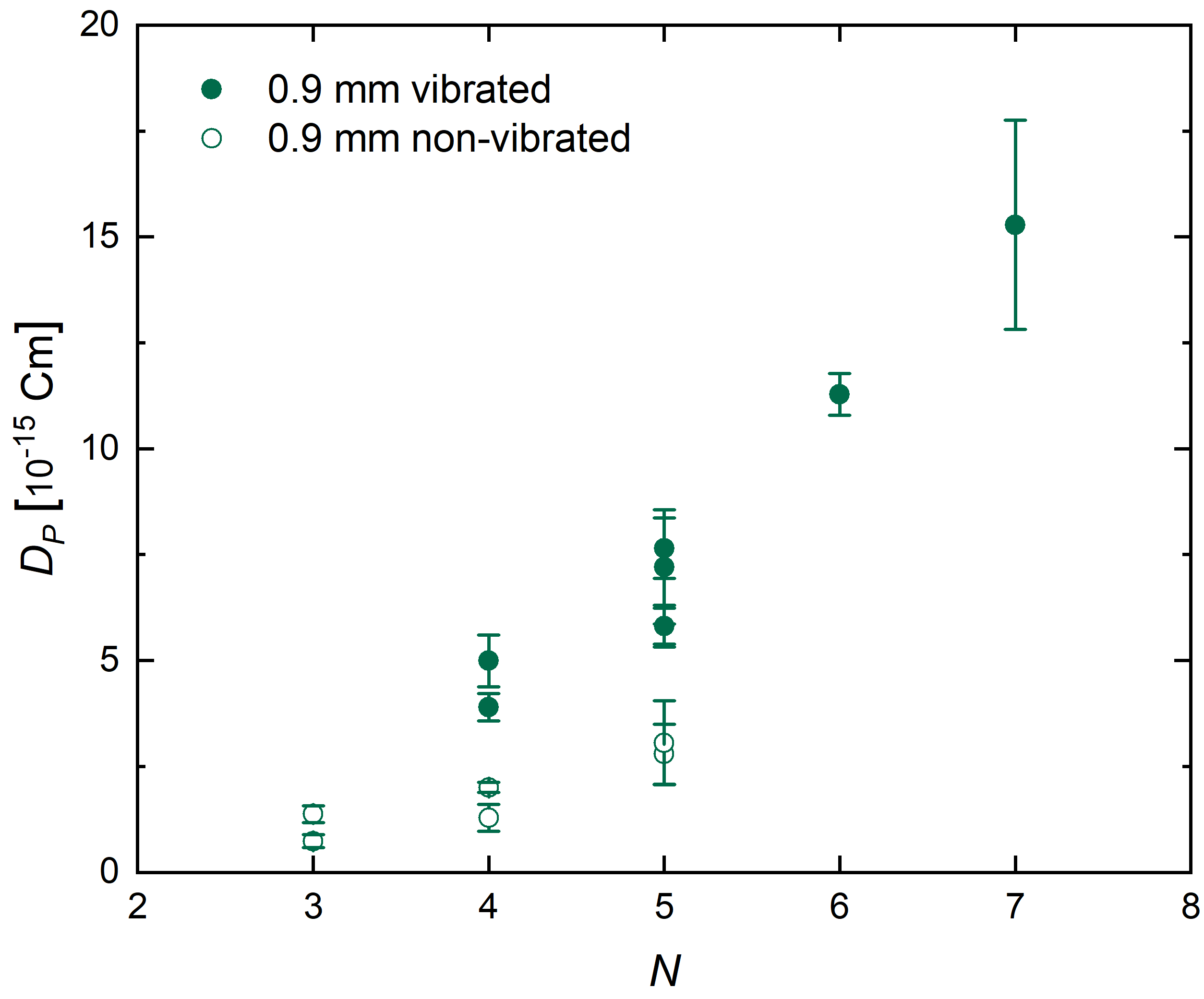}
	\caption{\label{dovern}Dipole moment dependence on cluster size $N$ for vibrated and non-vibrated samples. The error bars are standard errors.}
\end{figure}

The permanent dipole moments measured are given in table \ref{table}.
\begin{table}
\caption{\label{table}Measured values for vibrated (*) and non-vibrated basaltic samples of different cluster configurations. Marked in red are the individual grains that were traced in order to obtain the data.}
\begin{tabular}{ |c|c|c|c|c| } 
  \hline
 		$R$ [mm] & Cluster & $N$ & $D_P$ [10$^{-15}$ Cm] & $\alpha_P$ [°]\\ 
 		\hline
	0.9 *  & \includegraphics[height=0.5cm]{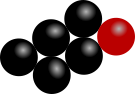} & 6 & 11.284 $\pm$ 0.489 & -4.630 $\pm$ 2.700\\ 
	& \includegraphics[height=0.5cm]{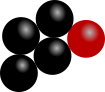} & 5 & 5.812 $\pm$ 0.494 & -5.140 $\pm$ 2.024\\
	& \includegraphics[height=0.5cm]{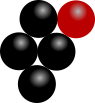} & 5 & 5.811 $\pm$ 0.429 & -8.875 $\pm$ 2.544\\
	& \includegraphics[height=0.5cm]{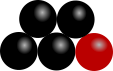} & 5 & 7.208 $\pm$ 1.350 & 2.810 $\pm$ -4.843\\
	& \includegraphics[height=0.5cm]{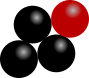} & 4 & 4.995 $\pm$ 0.611 & -0.830 $\pm$ 1.201\\
	& \includegraphics[height=0.5cm]{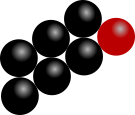} & 7 & 15.283 $\pm$ 2.467 & 2.308 $\pm$ 5.315\\
	& \includegraphics[height=0.5cm]{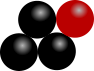} & 4 & 3.898 $\pm$ 0.326 & -7.633 $\pm$ 1.088\\
	& \includegraphics[height=0.5cm]{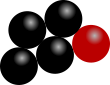} & 5 & 7.650 $\pm$ 0.716 & -14.694 $\pm$ 1.192\\
	  		\hline\hline
	0.9 & \includegraphics[height=0.5cm]{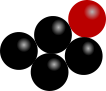} & 5 & 2.791 $\pm$ 0.710 & 25.223 $\pm$ 2.936\\
	& \includegraphics[height=0.5cm]{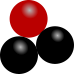} & 3 & 0.735 $\pm$ 0.153 & 135.786 $\pm$ 16.098\\
	& \includegraphics[height=0.5cm]{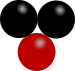} & 3 & 1.373 $\pm$ 0.204 & -80.553 $\pm$ 2.846\\
	& \includegraphics[height=0.5cm]{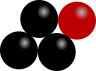} & 4 & 2.005 $\pm$ 0.121 & -17.401 $\pm$ 5.509\\
	& \includegraphics[height=0.5cm]{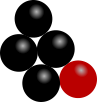} & 5 & 3.058 $\pm$ 0.989 & -4.853 $\pm$ 1.972\\
	& \includegraphics[height=0.5cm]{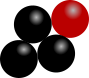} & 4 & 1.283 $\pm$ 0.317 & -20.091 $\pm$ 4.802\\
 		\hline\hline
	0.55 *  & \includegraphics[height=0.5cm]{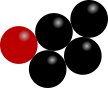} & 5 & 1.101 $\pm$ 0.052 & 206.519 $\pm$ 1.918\\
	& \includegraphics[height=0.5cm]{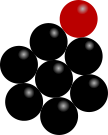} & 8 & 0.989 $\pm$ 0.836 & -13.178 $\pm$ 10.952\\
	& \includegraphics[height=0.5cm]{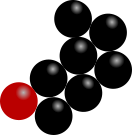} & 8 & 3.286 $\pm$ 4.601 & 141.652 $\pm$ 13.909\\
 		\hline\hline
	0.55 & \includegraphics[height=0.5cm]{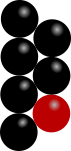} & 7 & 2.633 $\pm$ 0.083 & -152.594 $\pm$ 2.382\\
	& \includegraphics[height=0.5cm]{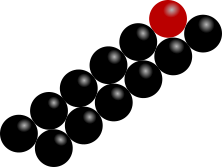} & 12 & 2.870 $\pm$ 0.206 & 3.158 $\pm$ 11.335\\
	& \includegraphics[height=0.5cm]{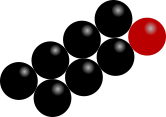} & 8 & 1.706 $\pm$ 0.217 & -33.272 $\pm$ 8.562\\
	& \includegraphics[height=0.5cm]{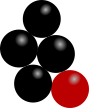} & 5 & 1.019 $\pm$ 0.082 & -142.112 $\pm$ 3.656\\
	& \includegraphics[height=0.5cm]{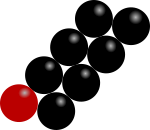} & 8 & 1.887 $\pm$ 0.205 & -212.490 $\pm$ 4.882\\
	& \includegraphics[height=0.5cm]{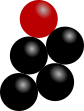} & 5 & 1.215 $\pm$ 0.122 & 216.399 $\pm$ 5.025\\
	& \includegraphics[height=0.5cm]{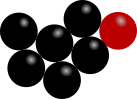} & 7 & 0.964 $\pm$ 0.129 & -11.694 $\pm$ 3.013\\
	& \includegraphics[height=0.5cm]{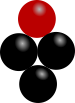} & 4 & 0.401 $\pm$ 0.085 & 32.304 $\pm$ 5.466\\
	& \includegraphics[height=0.5cm]{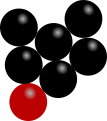} & 7 & 2.315 $\pm$ 0.180 & -171.945 $\pm$ 3.326\\
	& \includegraphics[height=0.5cm]{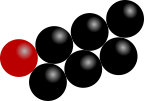} & 7 & 1.411 $\pm$ 0.197 & -2.809 $\pm$ 4.584\\
	& \includegraphics[height=0.5cm]{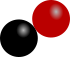} & 2 & 0.152 $\pm$ 0.071 & -6.759 $\pm$ 8.440\\
	& \includegraphics[height=0.5cm]{411} & 2 & 0.108 $\pm$ 0.022 & -25.777 $\pm$ 6.827\\
 \hline
\end{tabular}
\end{table}
The closest related work on dipole moments to which we can compare our results to are the ones measured by \citet{Steinpilz2020b}. They used spherical particles of about 0.9 mm, which is the same size as our large spheres and about twice as large as the small spheres. It is important to note, however, that they used glass spheres as sample material while we used basalt spheres. For dimers they find dipole moments between  $0.4 \cdot 10^{-16} \rm C m$ and $3 \cdot 10^{-16} \rm C m$. The only two dimers we measured for the smaller grains have dipole moments on the order of $1 \cdot 10^{-16} \rm C m$. If charge and therefore dipole moment scales with the surface area, our data points will fit to the upper end of the data range found by \citet{Steinpilz2020b}.  In addition, we have to consider that our dimers were not vibrated beforehand. Considering the difference in dipole moment between vibrated and non-vibrated samples given below (fig. \ref{dovern}), dipole moments on vibrated dimers will be larger than the values by \citet{Steinpilz2020b}, which were also precharged by collisions. This discrepancy can be explained by the material dependence regarding electric charging. We used glass and basalt spheres in different charging experiments in the past \citep{Jungmann2021conserve, Wurm2019, Steinpilz2020a, Teiser2021, Jungmann2018}. Basaltic samples typically exhibit higher net charges, but the experiments mentioned cannot easily be compared to each other. In any case, in otherwise similar experimental settings, basalt typically carries a net charge that is a factor of about 5 higher than charges found on glass particles. This value is based on microgravity experiments and is unpublished (Bila et al. in prep). Assuming that this scaling in net charge between the two materials also holds for the permanent dipole moments, our data is consistent to the data by \citet{Steinpilz2020b}. So we consider our methodology well suited to study dipole moments and well connected to the work by \citet{Steinpilz2020b}.


They considered different possible dipole and charge configurations in detail finding that such large dipole moments can only be explained by a patched charge model, and not by single net charges placed at extreme distances to each other. We did not measure the net charges on grains, yet. It is a logical extension for the future, but it is beyond the current work.
Anyhow, we now have a verified, robust and repeatable method that allows the determination of the permanent dipole moment of (sub)-mm particles. 

This allows parameter studies of various kinds. Similar to net charge, the dipole moment might depend on several variables including material, temperature, humidity and also charging history. Quantification and comparison with numerical calculations might help to constrain charging models further.

As first indication, fig. \ref{dovern} shows the dipole moment in dependence of the number of particles $N$ per cluster. There is a clear increase in dipole moment which for the vibrated sample seems to be linear. For a sample not vibrated before trapping the dipole moment is systematically smaller. For a given number of particles, different configurations of aggregates are possible. If there is a systematic dependence between charging, dipole moments and configurations it will be interesting to see. \citet{Lee2015}, for example, observed crystal-like structures in clusters with grains of different polarity. However, in view of identical grains being used here with low charging bias, this might be more subtle and cannot be studied with the given data set. More data are necessary for an interpretation but this already shows that collision history and the aggregates' size are major parameters to determine the initial dipole moment of an aggregate.

We have to emphasize "initial" as the dipole moment might change over time.
We already mentioned the timescales of charge redistribution, which now become accessible. On a conducting sphere, the charge should be homogenized. This will not change the net charge but will change the dipole moment. If the dielectric grains are not perfect insulators, homogenization might also occur. As grains can be levitated in an acoustic trap without a time limit and as the environment can be changed on purpose, the evolution of the dipole moment can trace this charge redistribution.
Along the same lines, a constant external electric field increases the dipole moment on slightly conducting grains over time and, therefore, also charge mobility on grains that are initially without dipole moment can be traced. 

The given data presented here are consistent with a constant permanent dipole moment over the time of the measurements. Therefore, they show that charge redistribution does not occur on timescales of minutes for basalt in a 1 bar air atmosphere as a first data point. 

\begin{figure}
    \centering
	\includegraphics[width=\columnwidth]{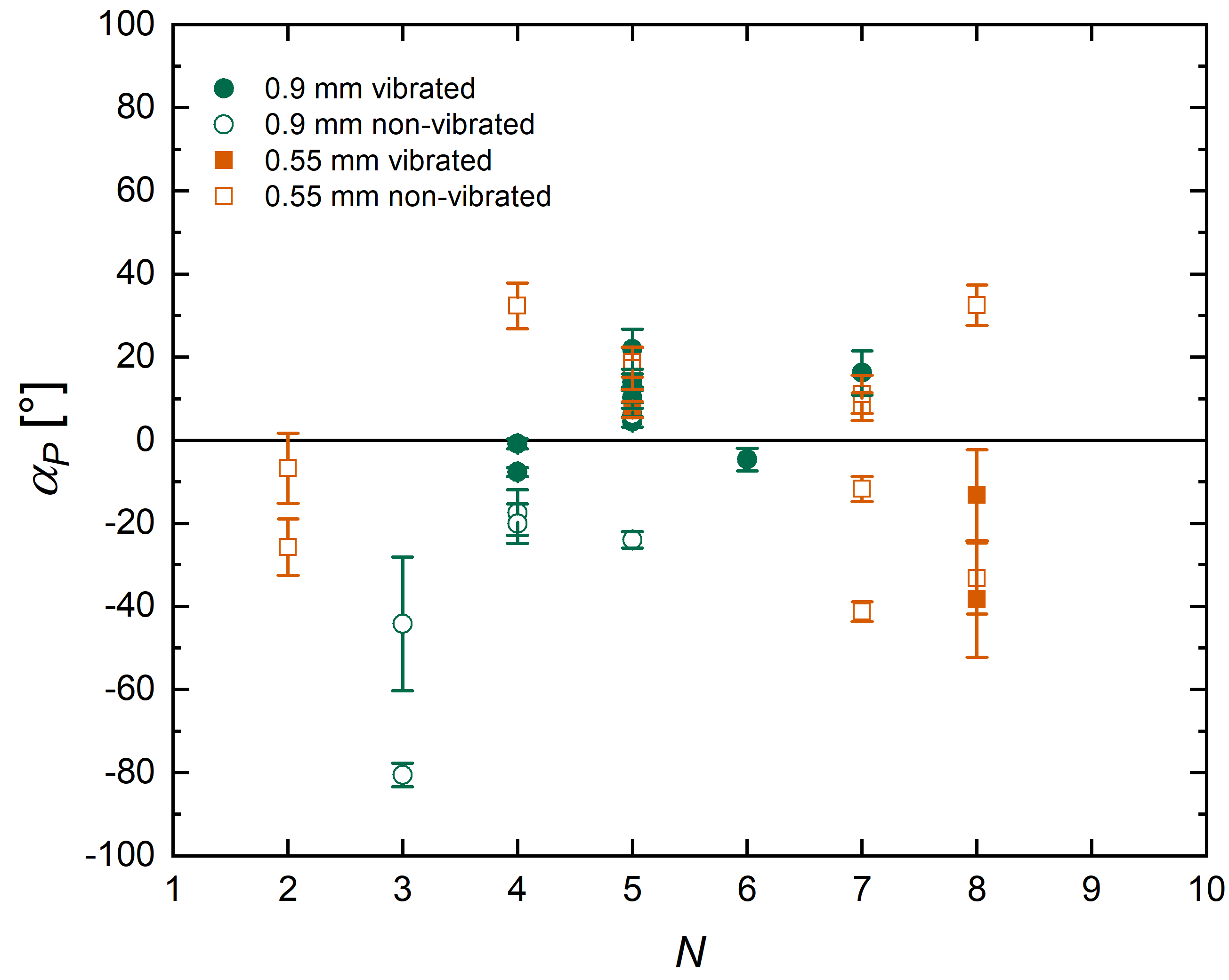}
	\caption{\label{angles}Electrical dipole moment alignment angle dependence on cluster size $N$ for all samples. The error bars are standard errors.}
\end{figure}

Above, we argued that we do not measure induced dipoles. The dependence of the measured dipole moments on the collisional history as seen in fig. \ref{dovern} clearly supports this view. Nevertheless, also for permanent dipoles, dipole moments in an elongated aggregate are likely directed along the long axis. As we track particles at the edge of the aggregates, the long axis is correlated to angles around 0 or 180$^{\circ}$.
Indeed, our data cluster at these angles as shown in fig. \ref{angles}. Here, for symmetry reasons for large absolute angles, 180$^{\circ}$ were added or subtracted. Besides, the angles from table \ref{table} were adapted regarding the longitudinal axis of the given cluster rather than the traced particle. Therefore, even aggregates with non-homogeneous charge distribution tend to align to their long axis in an electric field.

\section{Conclusion \label{conclusion}}

First and foremost, we outlined a method that allows the determination of permanent dipole moments on (sub)-mm grains levitated in an acoustic trap.
The first measurements on basaltic grains confirm that dipole moments are large on grains charged by collisions, which suggests that the dielectric surfaces initially have a patchy charge distribution.

However, since they are not in contact with other surfaces, this offers further possibilities, e.g. for studying the time dependence of a dipole moment without the influence of disturbance due to the placement or removal of a grain. Here, a first finding is that the basalt grains do not change their charge distribution on timescales of minutes as a lower limit. Clearly, this is a road laid out by this method for detailed mobility studies.

Furthermore, our first data show that dipole moments increase with aggregate size and that the collisional history is traceable by the dipole moment and that aggregates align with their long axis along electric fields, even without induced dipole moments. 

Overall, the acoustic trap is a suitable tool to investigate these effects and their dependencies on various parameters in detail.\\[1cm]

\noindent
\textbf{Acknowledgments}

\noindent
This project is funded by the Deutsche Forschungsgemeinschaft (DFG, German Research Foundation) – 458889524.
F. C. O. is supported by DLR Space Administration with funds provided by the Federal Ministry for Economic Affairs and Energy (BMWi) under grant numbers 50\,WM\,2142. We also thank the two referees for their constructive review. 

\bibliographystyle{mnras}
\bibliography{apssamp} 

\bsp	
\label{lastpage}
\end{document}